\documentclass[twocolumn,superscriptaddress]{revtex4}
\usepackage{graphicx}
\usepackage{epstopdf}
\usepackage{amsmath}
\usepackage{amssymb}
\usepackage{hyperref}
\usepackage{booktabs}
\usepackage{color}
\usepackage{setspace}

\usepackage{graphicx}

\usepackage{chapterbib}

\begin{document}

%
%
%
%

\title{Assessing the role of static lengthscales\\behind glassy dynamics in polydisperse hard disks}
\author{John Russo}
\email{russoj@iis.u-tokyo.ac.jp}
\author{Hajime Tanaka}
\email{tanaka@iis.u-tokyo.ac.jp}
\affiliation{Institute of Industrial Science, University of Tokyo, 4-6-1 Komaba, Meguro-ku, Tokyo 153-8505, Japan}


\begin{abstract}
The possible role of growing static order in the dynamical slowing down towards the glass transition
has recently attracted considerable attention.
On the basis of random first-order transition (RFOT) theory,
a new method to measure the static correlation length of amorphous order, called ``point-to-set (PTS)'' length,
has been proposed, and used to show that the dynamic length grows much faster than the static length.
Here we study the nature of the PTS length, using a polydisperse hard disk system,
which is a model that is known to exhibit a growing hexatic order upon densification.
We show that the PTS correlation length is
decoupled from the steeper increase of the correlation length of hexatic order,
while closely mirroring the decay length of two-body density correlations.
Our results thus provide a clear example that other forms of order can play an
important role in the slowing down of the dynamics,
casting a serious doubt on the order agnostic nature of the PTS length
and its relevance to slow dynamics, provided that a polydisperse hard disk system is a typical glass former.
\end{abstract}

\maketitle

\section*{Introduction}

When we supercool a liquid while avoiding crystallization, dynamics becomes heterogeneous~\cite{kob1997dynamical,yamamoto1998dynamics} and slows down significantly towards the glass transition,
below which a system becomes 
a non-ergodic state. Now there is a consensus that this slowing down accompanies the growth of dynamical correlation length~\cite{berthierR}. 
Several different physical scenarios have been proposed, yet the origin is still a matter of serious debate:
while some scenarios describe the glass transition as a purely kinetic phenomenon~\cite{chandler2009}, 
others posit a growing static order~\cite{royall2014role} or a loss of configurational entropy~\cite{starr2013relationship} behind dynamical slowing down.
Among this last category, we will focus here on two distinct approaches.
The first one is random first-order transition (RFOT) theory \cite{Kirkpatrick,lubchenko2007,Parisi2010},
which is based on a finite dimensional extension of mean-field models with an exponentially large number of metastable states.
The second approach, recently proposed by some of us~\cite{tanaka2012bond,tanaka2010critical},
ascribes the growth of the dynamical correlation length with the corresponding growth of the static correlation length.
Here we focus on these two scenarios based on static order and consider 
which is more relevant to the origin of glassy slow dynamics, using a simple model glassformer, two-dimensional (2D) polydisperse hard disks~\cite{KAT,kawasaki2011structural}. 

In RFOT theory, metastable states are thought to have amorphous order, whose correlation length diverges towards the ideal glass transition point. 
It was recently suggested that the so-called point-to-set (PTS) length, which is the correlation length of amorphous order,
can be extracted by pinning a finite fraction of particles and studying the dependence of the overlap function on the pinning particle concentration. 
According to the RFOT theory, amorphous order develops in any glass-forming liquids and this method is thought
to be able to pick up the static correlation length whatever the order is, i.e., the method is claimed to be order agnostic~\cite{charbonneau_static}. 
Thus, the use of pinning fields has been considered to be a promising new direction
in the study of the glass transition. Within the RFOT theory, it was shown that
freezing the positions of a finite concentration of particles shifts the ideal glass transition to higher
temperatures, potentially granting access to the glass state in equilibrium~\cite{CavagnaR,Cavagna2012,cammarota2012ideal,cammarota2013random,kob2013probing,karmakar2013random,chakrabarty2014phase,ozawa2014equilibrium}.
Moreover, the average distance between pinned particles at the liquid-to-glass transition represents a direct measure
of the PTS correlation length. PTS correlation lengths aim at measuring hidden static length scales
by looking at the extent of the perturbation induced by frozen particles on the rest of the liquid.
It is intuitively defined as the average distance between pinned particles that forces the system to stay in
an amorphous configuration with a vanishing configurational entropy. The reasons behind the popularity
of PTS correlation lengths in the study of the glass transition are at least twofold: 1) they are expected to 
provide an ``order agnostic'' method to measure static correlations~\cite{charbonneau2013decorrelation,PhysRevLett.113.157801}; 2) it is theoretically established
that no divergence of the relaxation time of a glass at finite temperature can occur without the concomitant
divergence of the static correlation length~\cite{montanari2006rigorous}.

On the other hand, 
it was recently noted~\cite{tanaka2010critical,kawasaki2011structural,tanaka2012bond} that, for moderately polydisperse hard disks,
an increase in the area fraction of particles $\phi$, hexatic (or 6-fold bond orientational) 
order grows and its correlation length $\xi_6$ is supposed to diverge, obeying the Ising-like power law, towards 
the ideal glass transition point $\phi_0$, where the structural relaxation time $\tau_\alpha$ diverges following the Vogel-Fulcher-Tammann law.  
We have also confirmed that the dynamical correlation length $\xi_4$ is proportional to the hexatic correlation length $\xi_6$ and furthermore there is almost 
a one-to-one correspondence between the degree of hexatic order and the slowness of dynamics. These results suggest an intimate link between static order and dynamics: 
the dynamical slowing down is accompanied by an increase in both size and
lifetime of hexatic ordered regions. 
We also found that 3D polydisperse hard and Lennard-Jones spheres exhibits essentially the same behaviour~\cite{tanaka2010critical}.
These results suggested that the dynamical slowing down is a consequence of the growing activation energy 
associated with the Ising-type power-law growth of the correlation length of critical-like fluctuations of static order towards the ideal glass transition point  
\cite{tanaka2012bond,tanaka2010critical}. Recently, a theory for the occurrence of such criticality in disordered
systems with topologically ordered cluster of particles was also proposed by Langer~\cite{langer}. 

The role of local order on the dynamics of glassy systems remains controversial at least for two reasons.
The first problem is that the local order that one needs to measure is system-dependent, and up to now
the relevance of bond orientational order was demonstrated only for polydisperse particle systems and a spin liquid, and not for bidisperse systems~\cite{tanaka2012bond,tanaka2010critical}. 
The second problem is conceptual: are static correlations really responsible for the dynamical slowing down?
The PTS correlation length is often described as a remedy to both problems, since it should be able
to detect static correlations without a detailed knowledge of the local order involved in these correlations.
In studies of binary mixtures of hard spheres, the PTS correlation length was shown to grow only
modestly in the regime accessible to computer simulation~\cite{PhysRevLett.108.035701,charbonneau2013decorrelation},
differently from the dynamical correlation length which grows much more rapidly. These results
suggested that no link exists between a single static length scale and the dynamical slowing down
(the only exception would be close to a possible ideal glass transition temperature)~\cite{charbonneau_static,royall2014role}. On the other hand,
measures of the PTS correlation length~\cite{reichman} have shown that it correlates
well with the average dynamics of the system, and with dynamic heterogeneities~\cite{flenner2013universal},
The PTS correlation length is in agreement with measures of the density of plastic modes~\cite{procaccia},
providing support for the idea of a fundamental length scale controlling the dynamics of the supercooled liquids.

\begin{figure*}[!t]
 \centering
 \includegraphics[width=12cm]{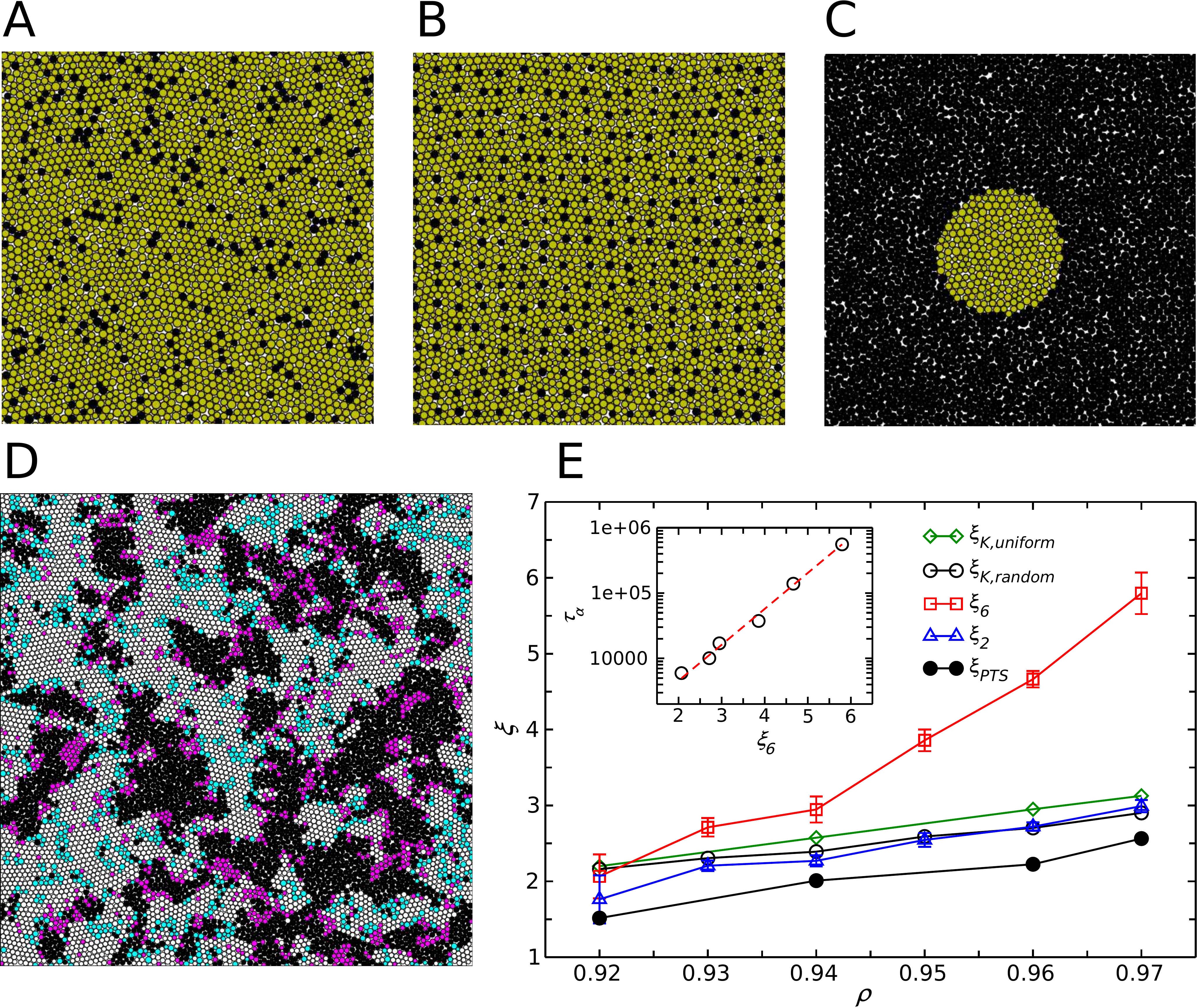}
 \caption{{\bf Types of pinning fields and the growth of static and dynamic correlation lengths.}
 (top row) Different pinning strategies, with pinned particles coloured in black: random pinning ({\bf a}), uniform pinning ({\bf b}) and cavity pinning ({\bf c}).
 {\bf d,} Snapshot of a configuration at $\rho=0.97$ in which
 the disks are coloured according to the following criteria: \emph{White}, low mobility and high order;
 \emph{Black}, high mobility and low order; \emph{Cyan}, low mobility and low order; \emph{Magenta} high mobility
 and high order.
 {\bf e,} Correlation lengths as a function of density $\rho$: red squares for bond orientational order $\xi_6$, blue triangles
for the two-body correlations $\xi_2$, black circles $\xi_K$ for random pinning, green diamonds for
$\xi_K$ for uniform pinning, black filled circles for $\xi_\text{PTS}$. The inset shows the scaling of 
the structural relaxation time $\tau_\alpha$ with the hexatic correlation length $\xi_6$ (points), and 
the fit with the relation $\tau_\alpha=\tau_0\exp(D\xi/\xi_0)$ (dashed line).}
 \label{fig:lengthscales}
\end{figure*}

Unlike previous studies, in this work we measure the PTS correlation length
in a system for which a growing local order was previously found, i.e., polydisperse
hard disks. This will allow us to compare 
the growth of PTS correlation lengths with the correlation length of bond orientational order.
We will consider several pinning strategies (random pinning, uniform pinning and cavity pinning) (Fig.~\ref{fig:lengthscales}\emph{A}-\emph{C}), and then look for the underlying structural
features that are captured by the PTS correlation length.
In principle, each different pinning geometry probes a different lengthscale~\cite{berthier2012static}. The \emph{point-to-set}
static lengthscale $\xi_\text{PTS}$ was first introduced in the spherical cavity geometry~\cite{biroli2008thermodynamic}.
Both random and uniform pinning are expected to express the same static lengthscale, here called $\xi_K$, and
the RFOT theory predicts a temperature scaling relation between $\xi_\text{PTS}$ and $\xi_K$, which in its simplest
form is written as $\xi_K(T)\sim \xi_{PTS}(T)^{1/2}$~\cite{cammarota2012ideal}.
If the PTS correlation method is indeed order agnostic, it should be able to pick up the correlation length of hexatic order in a polydisperse hard disk system, 
provided that it is a typical glass former. 
Thus, it is a main interest of this work to reveal whether the PTS length is the same as the hexatic correlation length. 
This question is of crucial importance to reveal the origin of slow glassy dynamics.

\section{Results: unpinned case}
The system studied is composed of $N=10000$ polydisperse hard disks with disk-size polydispersity $\Delta=11\%$ (see Methods). We start by considering the case without
an external pinning field, $c=0$. We focus on the following number densities from $\rho=0.92$ to $\rho=0.97$,
which correspond to area fractions ranging from $\phi=0.73$ to $\phi=0.77$.
As described in {\it Supplementary Information}, we extract the correlation length for bond-orientationally ordered
regions by fitting the exponential decay of the peaks of the correlation function  $g_6(r)/g(r)$ (see Fig. S1). 
We use an exponential function
instead of a 2D Ornstein-Zernike function to avoid \emph{a priori} assumptions on the origin of the growth of the correlation
length. The correlation length $\xi_6$ is plotted in Fig.~\ref{fig:lengthscales}\emph{E}, together with other length scales which we will
derive later.
The two-body correlation function $\xi_2$ is obtained by fitting with an exponential law the decay of $g(r)-1$. The results of
this fit are also summarized in Fig.~\ref{fig:lengthscales}\emph{E}. These results confirm that, for polydisperse
glass forming systems, the growth of many-body correlations associated with bond orientational order is much faster than
the growth of two-body correlations~\cite{tanaka2010critical,kawasaki2011structural,leocmach2013importance}.

As was shown in Refs.~\cite{tanaka2010critical,kawasaki2011structural} there is a link between dynamic heterogeneities
and regions of high hexatic order in the fluid. In Fig.~\ref{fig:lengthscales}\emph{D} we show the results of Event-Driven
molecular dynamics simulations in the isoconfigurational ensemble (see Methods), where $200$ trajectories
are started from the same initial configuration but with different initial velocities. The degree of structural order is investigated
by taking the average hexatic field over these $N=200$ trajectories (also called isoconfigurational average) at
a time $t=\tau_\alpha/10$, where $\tau_\alpha$ is the structural relaxation time measured through the intermediate
scattering function. The dynamics is instead investigated through the isoconfigurational average of the
relative displacement $|\mathbf{R}_i|$ after $t=\tau_\alpha$, which is also approximately the time at which the heterogeneities are
maximum (as measured by the four-point susceptibility~\cite{berthierR}). The relative displacement $\mathbf{R}_i$ is defined
as the displacement between time $t=0$ and $t=\tau_\alpha$, $\mathbf{r}_i(\tau_\alpha)-\mathbf{r}_i(0)$, of particle $i$ with respect to its $M$ neighbours, $\mathbf{R}_i=\mathbf{r}_i-\frac{1}{M}\sum_j^M \mathbf{r}_j$.  
This operation is introduced since coherent translational motion of a group of particles does not contribute to the stress relaxation.
All disks in the configuration of Fig.~\ref{fig:lengthscales}\emph{D} 
are then grouped into sets of high and low mobility/order depending
whether their mobility/order is higher or lower than the 50th percentile. We can then identify four different sets of particles:
low mobility and high order (white); high mobility and low order (black); low mobility and low order (cyan); high mobility
and high order (magenta). Our results show that 76\% of particles are in the first two sets (38\% in each), demonstrating
a high degree of correlation between structural ordered regions and immobile regions (or, vice versa, between
disordered regions and mobile regions). Moreover, the remaining two sets (each accounting for the 12\% of particles)
are located at the interface between mobile and immobile extended regions. In Fig.~\ref{fig:lengthscales}\emph{E} magenta
disks are located on the surface and in between black clusters, while cyan disks are located on the surface and
in between white clusters. In other words disks which are next to
an low/high mobility region, will also have low/high mobility. 
Here we note that the embedded fractal nature of order parameter fluctuations is characteristic of critical fluctuations. 
 
\section{Results: pinning}
Having characterized the static properties of the unperturbed system, we now introduce the pinning field.
First, simulations are fully equilibrated with cluster-moves algorithms (see Methods).
A representation of the pinning fields is given in Fig.~\ref{fig:lengthscales}\emph{A}-~\emph{C}. 
In the \emph{random pinning} geometry, $N_p$ particles are chosen
randomly and pinned, i.e. their position is kept fixed during the course of the simulations.
For each density we introduce pinning fields with concentrations
$c=0.01,0.06,0.10,0.15,0.20$, and for each concentration we average over 9 different realizations of the fields.
In this scheme the distance between pinned particles is defined only as an average over a broad distribution, as both clusters of pinned particles and extended regions without pinned particles
are likely produced. For this reason, random pinning is expected to be more sensible to finite size effects,
as was observed in Ref.~\cite{kob2013probing}, where it is noted that random pinning can smear out the Kauzmann 
transition in very small systems. The pinning geometry can also have strong effects on the dynamics~\cite{berthier2012static,jack2013dynamical,chakrabarty2014phase,kob2014non}.
In order to limit the fluctuations in the distance between pinned particles, we also adopt a \emph{uniform pinning} geometry,
where a simple cubic lattice is overlaid to the equilibrated configuration,
and the closest particle to each lattice point is pinned.
Particles are pinned at the following average distances: $a=2.5,2.75,3,4.25,6,8,10$, and each distance is
averaged over $7$ realizations of the field.
Finally, in the \emph{cavity pinning} geometry, all particles outside a cavity of radius $R$ are pinned. Since the static
 length scales currently accessible to simulations are expected to be small, well within $10\sigma$, simulations
 with cavity pinning involve a small number of particles, thus requiring extensive average over different realizations
 of the field (here $100$ simulations for each cavity diameter).

Since the pinning field is applied to equilibrium configurations, the static properties should be unchanged with respect
to the $c=0$ case. We check this by computing both positional and hexatic order for different concentrations $c$. All results
are consistent with the $c=0$ case and the standard deviation between simulations at different $c$ is represented with the error bars
for $\xi_6$ in Fig.~\ref{fig:lengthscales}.
Correlation lengths are extracted from all pinning geometries, following the procedure outlined in {\it Supplementary Information}. In all
cases the physical idea is to detect the characteristic length (average distance between pinned particles in the random and uniform
pinning geometries, or the size of the cavity in the cavity geometry), which produces a high localization of the mobile particles,
as measured by overlap functions.
We plot the random pinning correlation length $\xi_{K,\text{random}}$, the uniform pinning correlation length $\xi_{K,\text{uniform}}$, and the point-to-set
lengthscale from cavity pinning $\xi_\text{PTS}$ in Fig.~\ref{fig:lengthscales}\emph{E} (see {\it Supplementary Information} on the details of its estimation). 
We see that the growth of the PTS lengthscale, irrespective of the pinning strategy, is significantly slower than the
growth of bond orientational correlation length $\xi_6$, while being comparable to the growth of pair correlations in the system $\xi_2$.
We also confirm that estimating the PTS length from uniform and random pinning through the relation $\xi_{PTS}(T)\sim\xi^2_K(T)$
still produces a much weaker growth than that of $\xi_6$.
The inset of Fig.~\ref{fig:lengthscales}\emph{E} shows the scaling between the $\xi_6$ length scale and the relaxation time $\tau_\alpha$,
$\tau_\alpha=\tau_0\exp(D\xi/\xi_0)$, where $D$ is a measure of the fragility of the system~\cite{tanaka2012bond}. This scaling
also supports a direct connection between the growth of structural correlation and slow dynamics.

\begin{figure*}[!t]
 \centering
 \includegraphics[width=16cm]{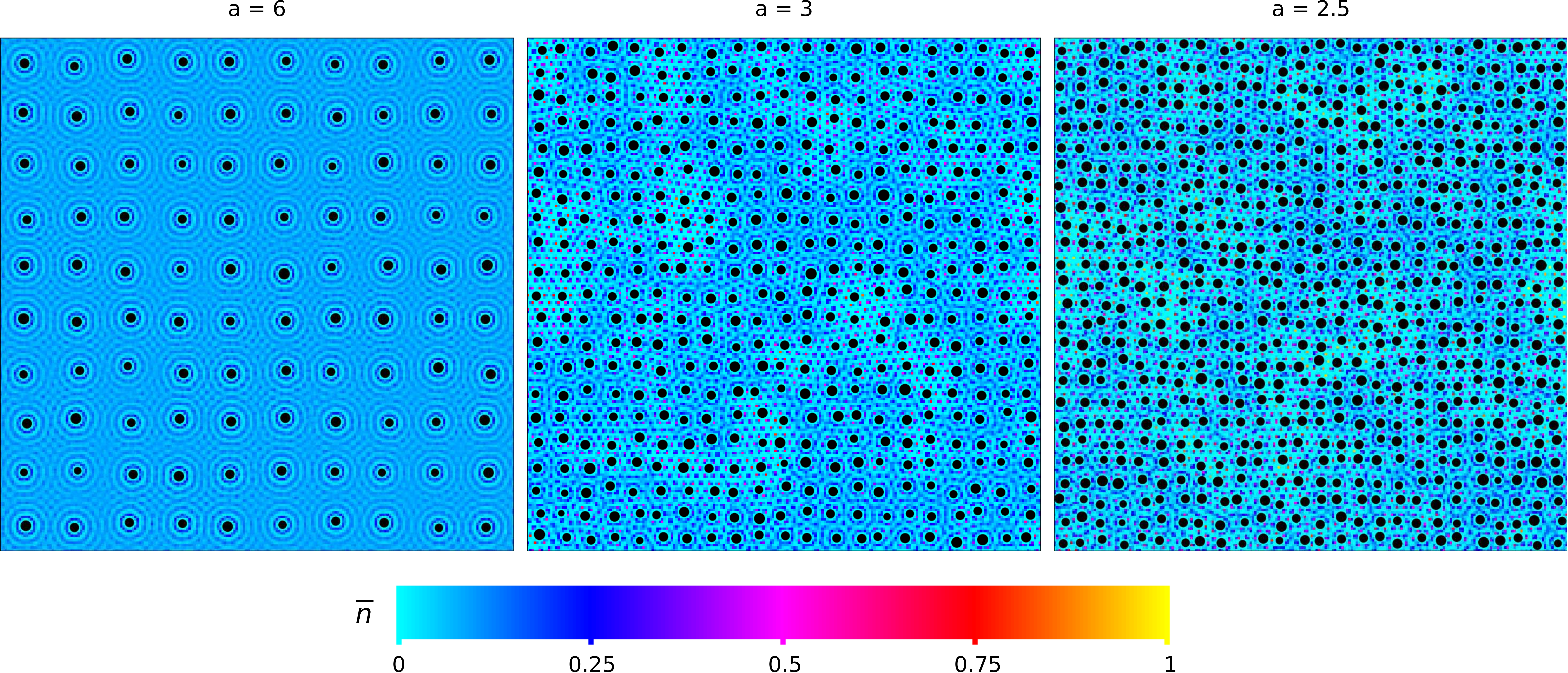}
 \caption{{\bf Development of $\overline{n_i}$ field with a decrease in $a$ for uniform pinning.} 
The particle density is $\rho=0.95$, and the average distance between particles
 $a=6$ (left), $a=3$ (middle) and $a=2.5$ (right). The colour code is mapped to the occupancy probability according
 to the colour bar. Distances are in units of the average diameter $\langle \sigma \rangle$.}
 \label{fig:uniform_presenceboxes}
\end{figure*}

All static lengthscales can also be obtained by considering coarse-grained variables, dividing the simulation box into smaller boxes of side length $l=0.3\sigma$, ensuring
that each box can be occupied at most by one disk at any time during the simulation. The problem is then mapped on a set of
discrete variables, defined as $n_i^\alpha=1$ when the center of a disk is in box $i$ in configuration $\alpha$, and $n_i^\alpha=0$
otherwise. The average occupancy for a particular realization of the pinning field is defined as $\overline{n_i}$,
where the overline denotes an average over thermal fluctuations for a fixed realization of the pinning field.
Figure ~\ref{fig:uniform_presenceboxes} shows the field $\overline{n_i}$ for three different concentrations of the pinned particles 
for the case of uniform pinning at $\rho=0.95$. Going from the low to high concentration (left to right in the figure) the amount of localization
in the field progressively increases. Localization is associated with regions with high occupancy probability, such as the purple dots
that can be seen for average distances $a=3\sigma$ and $a=2.5\sigma$. At $a=3\sigma$, these high occupancy regions are localized in
particular regions of the simulation box, while at $a=2.5\sigma$ they fill the box rather uniformly. This suggests that the PTS
correlation length, which characterizes the crossover of the localization transition, is between $2.5\sigma\lesssim\xi_{\rm PTS}\lesssim3\sigma$,
which is consistent with the measurement based on the spatial correlation of the overlap function, $\xi_{\rm PTS}({\rho=0.95})\approx 2.6\sigma$ 
(see Fig.~\ref{fig:lengthscales}\emph{E}). Moreover we confirm that all static lengthscales obtained through the coarse grained representation are fully consistent
with those obtained in Fig.~\ref{fig:lengthscales}\emph{E} (see {\it Supplementary Information}).
The similar scaling behaviour between pinning lengthscale $\xi_K$ and the pair correlation length $\xi_2$ can also be
understood in terms of the $\overline{n_i}$ field.
Each pinned particle generates an oscillatory perturbation of the $\overline{n_i}$ field,
which originates from the two-body static correlations between the pinned particle and the mobile particles in the liquid.
So no localization transition can occur if the average distance between pinned particles is bigger than the range of
two-body correlations, $\xi_2$. The localization transition thus requires pinned particles to be within $\xi_2$,
below which the number of particle arrangements drastically decreased,
and the configurational entropy vanishes. The extent of these regions with high localization of particles is exactly what
is being measured by the PTS correlation function.
 
The results thus show that the growth of the pinning correlation length is
similar to the growth of two-body correlations. On the other hand, the growth of the bond orientational order correlation length is much faster, and clearly
decoupled from the pinning correlation length. We also checked that the same is true for coarse-grained quantities.
This strongly indicates that the PTS correlation length is not \emph{order agnostic}, but targets the growth of a particular order in the system, that is
the size of the regions where particles are localized due to the pinning field. The growth of these regions follows the growth
of two-body correlations: particles are localized due to the perturbation that pinned
particles introduce to the $\overline{n_i}$ field, and the length scale of this perturbation is given by two-body correlations.
In other words, at least in the density range considered here,
the localization transition due to point pinning requires that the average distance between pinned particles is smaller than the two-body
correlation length.

\section{Results: localization}
\begin{figure*}[!t]
 \centering
 \includegraphics[width=16.cm,clip]{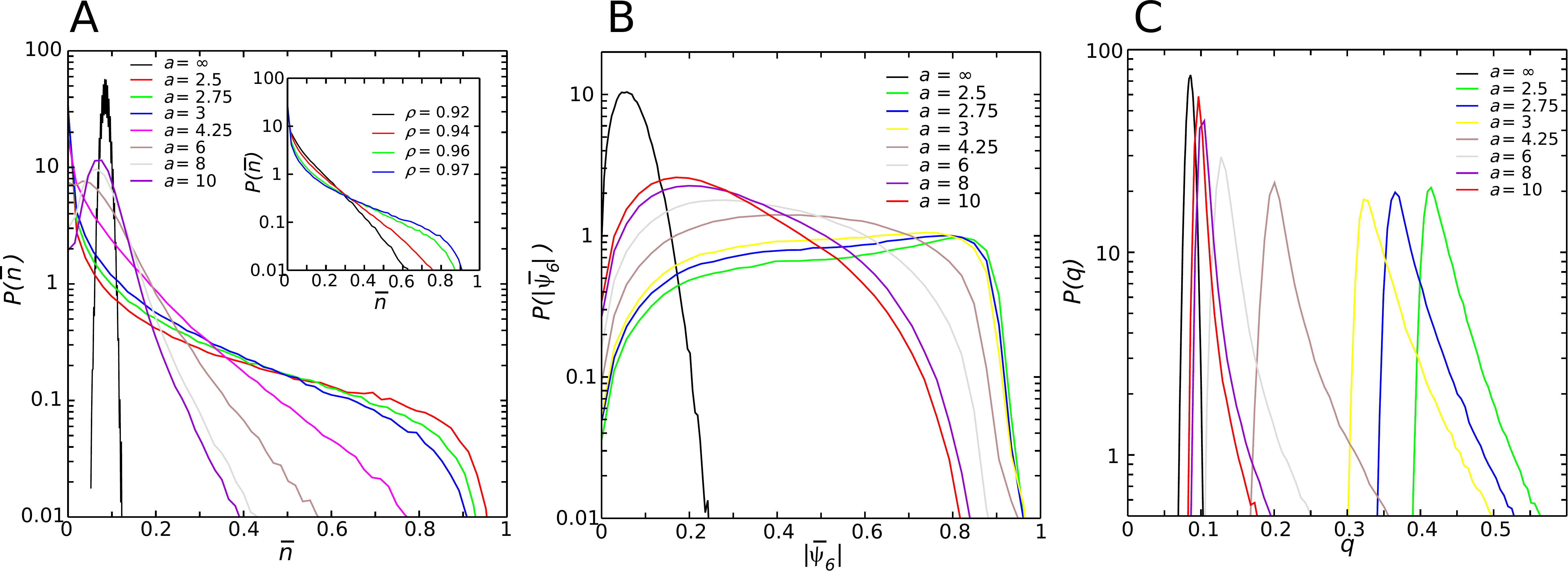}
 \caption{{\bf Dependence of the probability distribution functions of various fields on the average distance between pinned particles $a$ 
 for uniform pinning.}
 {\bf a,} Probability distribution function of the occupancy field $\overline{n_i}$ at $\rho=0.95$
 and different values of $a$. The inset shows $P(\overline{n})$ for constant $a=3\sigma$ and different
 values of $\rho$. {\bf b,} Probability distribution function for the average hexatic field magnitude, $\overline{|\psi_6|}$,
 at $\rho=0.95$ and for different values of $a$. {\bf c,} Probability distribution of the overlap, $P(q)$, for $\rho=0.95$ and for different values of $a$.}
 \label{fig:uniform_histograms}
\end{figure*}

Next we consider whether the localization transition is linked with the underlying hexatic ordering.
A visual inspection of the
configuration at $a=3$ in Fig.~\ref{fig:uniform_presenceboxes}, shows that the localized regions tend to form rather compact domains.
These domains are found by plotting the probability distribution function for the occupancy field $\overline{n_i}$, as shown in Fig.~\ref{fig:uniform_histograms}\emph{A}.
The figure plots $P(\overline{n})$ for different values of $a$, the average distance between pinned particles, at density $\rho=0.97$.
Without pinned particles ($a=\infty$) the distribution shows one Gaussian peak centred around $\overline{n}=l^2\rho$, where $l=0.3\sigma$ is the coarse-graining length.
For finite values of $a$, the distribution progressively broadens, and, for $a\lesssim 3$, covers almost all the $\overline{n}$ range, $\overline{n}\in [0;1]$.  
A similar behaviour is also seen for the distribution function of the coarse grained hexatic field $\overline{\psi_6}$, as shown in  
Fig.~\ref{fig:uniform_histograms}\emph{B}, where the distribution of $\overline{\psi_6}$ is plotted for the same state points of the top panel. 
A visual inspection of the underlying configurations indeed confirms that localized particles appear in regions where the average hexatic field is strong,
thus revealing an important structural feature of the localization transition. Localization appears in regions of high hexatic order, where the
average distance between pinned particles is within the two-body correlation length, $\xi_2$.

Now we examine whether the localization observed in our simulations is a true first order transition or just a crossover.
Within RFOT theory, the lower critical dimension is two, so a true thermodynamic transition is not
expected in our simulations, even though the question still remains open. Both distribution functions in Fig.~\ref{fig:uniform_histograms}\emph{A} and \emph{B} 
do not show signs of bimodality.
We next examine the global fluctuations
of the overlap function, $P(q)=\langle\delta(q-q_{\alpha\beta})\rangle$ (see Fig.~\ref{fig:uniform_histograms}\emph{C} for $\rho=0.97$). The average value of $q$ changes
continuously as a function of $a$, and no sign of bimodality is present. For small values of $a$ and high $q$ the distributions deviate
significantly from the Gaussian shape and display a heavy tail. This is not associated with the glass-transition localization but to a
different localization process which happens when the distance between two pinned particles is small enough to block the passage of fluid
particles. This is analogous to the mechanism responsible for the Lorentz-gas transition~\cite{horbach}.
We thus find no evidence of a first-order localization transition. Our results of course do not exclude the possibility that such
transition could occur at higher densities. The broad distribution observed in Fig.~\ref{fig:uniform_histograms}\emph{A} and \emph{B} are in fact compatible
with the presence of two population in a mixed state. The inset of Fig.~\ref{fig:uniform_histograms}\emph{A} plots the probability distribution function is plotted for $a=3\sigma$ and for different values of $\rho$: the different curves cross at an isosbestic point, which is a common feature of bimodal distribution functions, meaning that the overall population becomes more localized as $\rho$ is increased (the presence of the isosbestic point is independent of the value of $a$).
But simulations at higher volume fractions require considerablely larger system sizes (due to the rapid increase of the hexatic order correlation
length, $\xi_6$), and are outside the scope of the present investigation.

\section{Discussion and Conclusions}

We have extracted several static length scales from systems of polydisperse hard disks with polydispersity $\Delta=0.11$, in the range
$\rho\in [0.92;0.97]$.
The results confirmed that the length scale associated with bond orientational order
grows more rapidly than the length of pair correlations~\cite{tanaka2010critical,kawasaki2011structural,leocmach2013importance}.
The use of pinning fields enabled the calculation of the PTS correlation length, showing that it
grows only moderately with increasing supercooling, a result which is in agreement with measures of the PTS length in
binary mixtures~\cite{charbonneau2013decorrelation}. For polydisperse systems, the PTS correlation length is not coupled to that of bond orientational order, 
which is directly linked to the dynamical correlation length \cite{tanaka2012bond,tanaka2010critical}: 
the growth of the former is considerably slower than the latter.
For different glass forming systems, this suggests that also other forms of order originating from many-body interactions could
go undetected by PTS measures. 

The PTS length captures a localization transition that occurs in presence of pinned particles. This localization transition originates when the occupancy field, $\overline{n}$,
has extended regions of high probability due to neighbouring pinned particles. The perturbation that a single pinned particle produces in the $\overline{n}$ field
is due to \emph{pair correlations}. In absence of strong nonlinear effects, the length scale of the localized regions extends no further than two-body correlations.
This is the case in the density interval accessible to our simulations, where pinned particles
need to be placed closer than the pair correlation length in order to produce localized regions in the fluid.
A second requirement that our results suggest is that pinned particles should be in positions compatible with high local hexatic order.


The localization transition that occurs with increasing concentration of pinned particles happens continuously 
in the density range we could access in equilibrium. 
The results do not rule out the possibility that strong non-linearities will produce a localization
transition that extends beyond pair correlations for higher (but yet unreachable) densities.

To summarize, the PTS length measured by particle pinning simply reflects pair correlation and fails in detecting the correlation of bond orientational order (more precisely, 
hexatic order), which intrinsically originates from many-body interactions. Although the PTS length is decoupled from the dynamical correlation length, 
the hexatic order correlation is strongly coupled to it. This implies that slow dynamics in our system is controlled by the development of hexatic ordering, 
and not by translational order detected by the PTS correlation. 
Although the generality of this conclusion needs to be checked carefully, our study suggests that the PTS length is not order agnostic and 
the growth of the PTS length is not responsible for glassy slow dynamics at least for our system.

\vspace{1cm}
\noindent
{\bf Acknowledgements}

\noindent
We are grateful for valuable comments and constructive criticisms to Ludvic Berthier, Gulio Biroli, Patrick Charbonneau,Walter Kob, Jim Langer, David Reichman, Gilles Tarjus, and Sho Yaida. 
This study was partly supported by Grants-in-Aid for Scientific Research (S) and Specially Promoted Research from 
the Japan Society for the Promotion of Science (JSPS).


%
%
%

\centerline{\bf \large Supplementary Information}

\vspace{0.5cm}

\subsection*{Simulations}
We study two-dimensional polydisperse hard disks with Monte Carlo simulations.
The diameter $\sigma$ of the disks is extracted from a Gaussian distribution,
and the polydispersity is defined as the standard deviation of the distribution,
$\Delta=\sqrt{\langle\sigma^2\rangle-\langle\sigma\rangle^2}/\langle\sigma\rangle$.
In the present work we fix $\Delta=11\%$ for which no transition to an hexatic phase is observed.
The unit of length is set by the average disk diameter $\langle\sigma\rangle$.

All simulations are run at fixed densities $\rho=N/V$, with $N=10000$, with the event-chain algorithm~\cite{PhysRevE.80.056704,PhysRevLett.107.155704},
which allows for a fast equilibration even at very high densities. After the equilibration run,
we activate the pinning field and switch to Metropolis dynamics with swap moves between
randomly selected pairs of non-pinned particles.

The connection between static and dynamic lengthscale shown in Fig.~1\emph{D} was obtained with Event Driven simulations~\cite{bannerman2011dynamo}
in the isoconfigurational ensemble~\cite{widmer2004reproducible}, where $200$ trajectories are started from an equilibrated
configuration at $\rho=0.97$ but with a different assignment of initial velocities.

\subsection*{Estimation of $\xi_6$ and $\xi_2$}

\begin{figure}[!b]
 \centering
 \includegraphics[width=8cm]{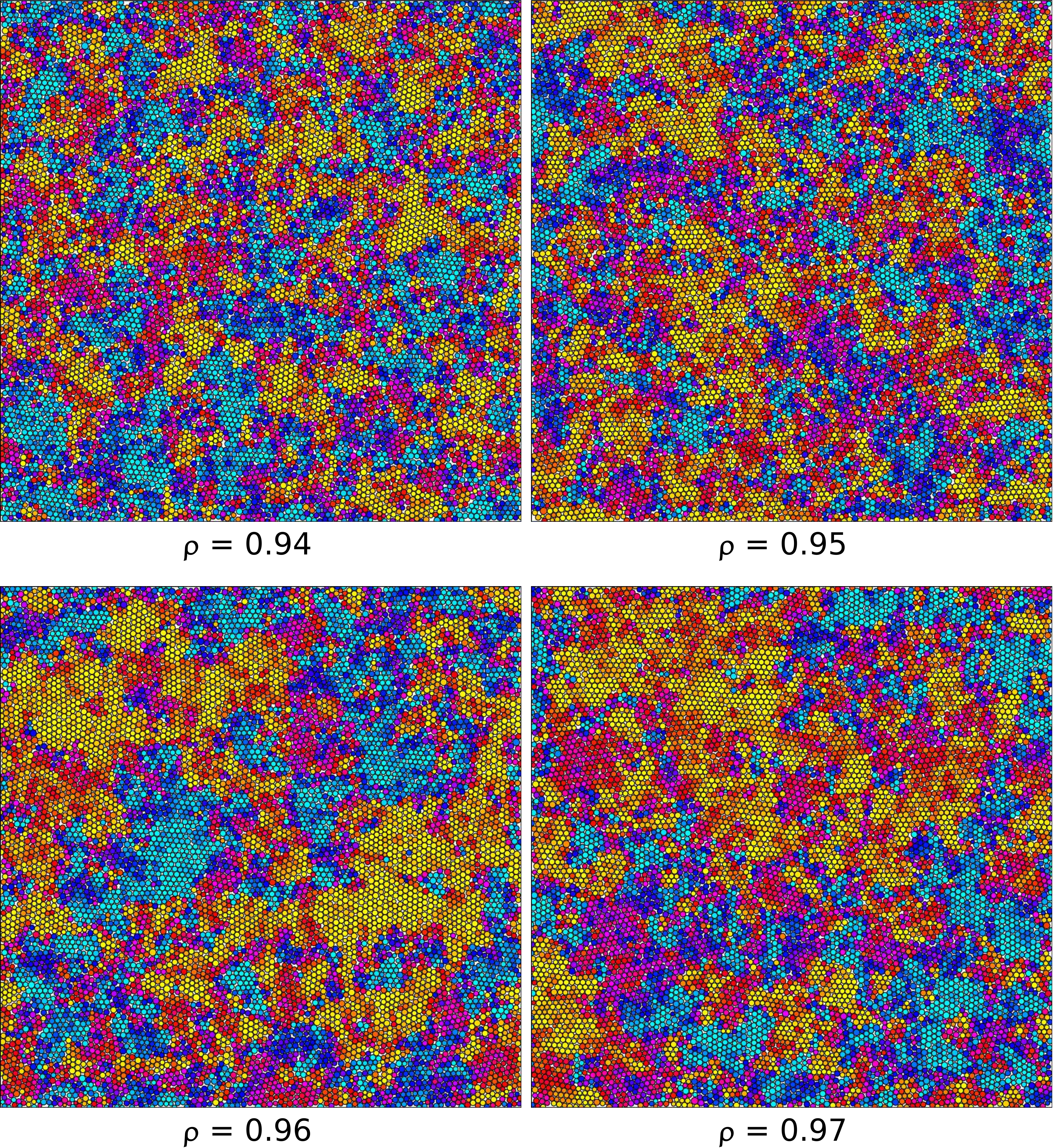}
 \caption{{\bf Snapshots of configurations of polydisperse hard disks at different densities.}
 Particles are coloured according to the phase of the hexatic order, with similar colours sharing the same orientation of the
hexatic order. The size of the correlated regions increases with density.}
 \label{fig:hexaticorder}
\end{figure}

Here we explain how to estimate the hexatic order correlation length $\xi_6$ and the translational correlation length $\xi_2$.
Figure~\ref{fig:hexaticorder} shows snapshots of configurations at different densities, where each particle is coloured according
to the phase of the local hexatic order, $\arg(\psi_6)$. The figures show that the hexatic order increases with $\rho$.

Bond orientational order in two dimensional polydisperse hard disks is expressed by the hexatic order parameter
$$
\psi_{6}=\frac{1}{n_j}\sum_ke^{i6\theta_{jk}}
$$
where $n_j$ are the neighbours of particle $j$, and $\theta_{jk}$ is the angle that the bond between particle $j$ and $k$ makes with a
reference axis. In the above definition, neighbouring particles are defined as particles sharing an edge in the radical Voronoi diagram
obtained from the particle's positions and sizes. To measure the extent of hexatic order we define the following correlation function
$$
g_6(r)=\langle\psi_6^*(r)\psi_6(0)\rangle
$$
and extract the correlation length from the spatial decay of $g_6(r)/g(r)$.

\begin{figure}[!t]
 \centering
 \includegraphics[width=8cm]{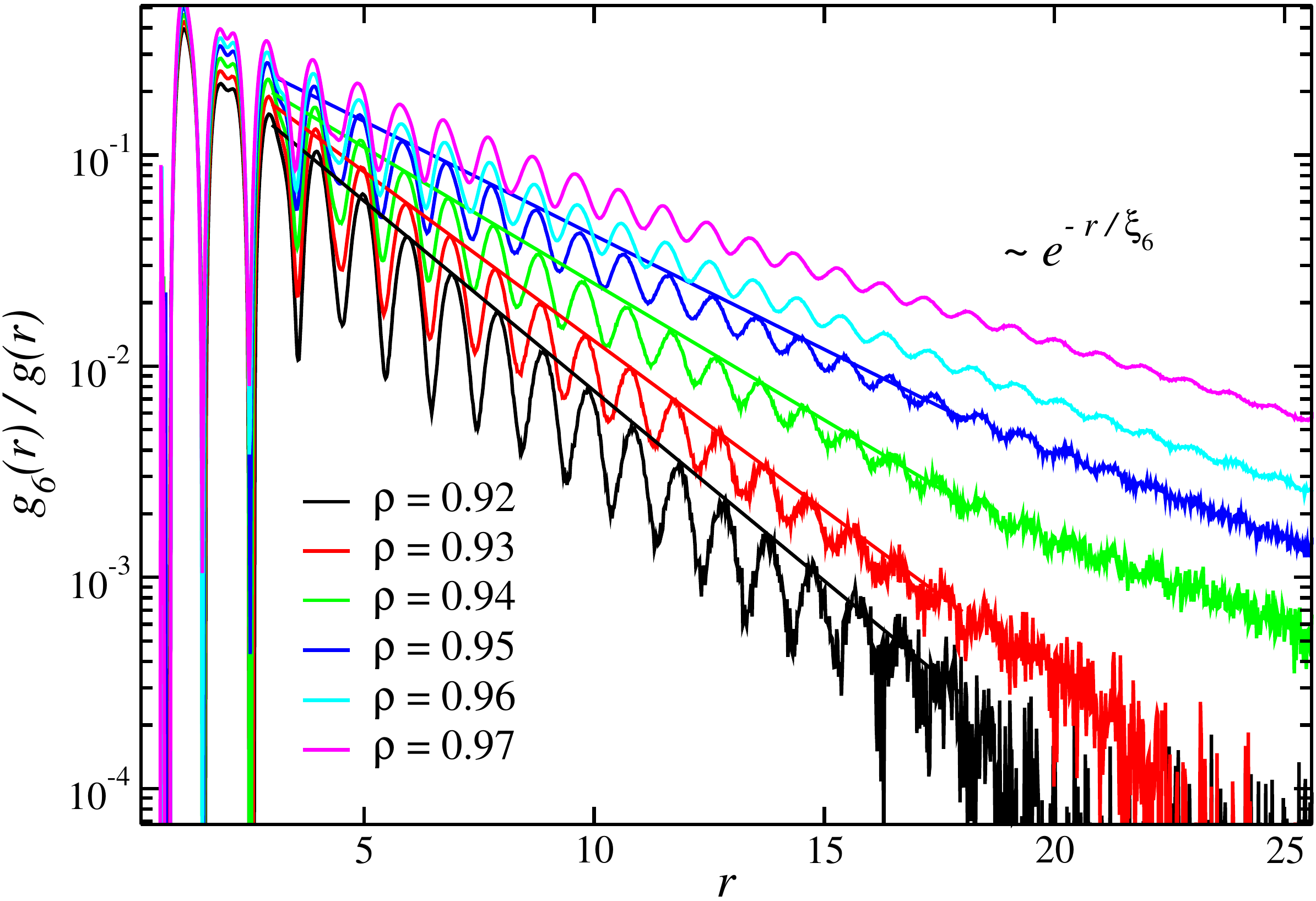}
 \caption{{\bf Decay of the hexatic order correlation function $g_6(r)/g(r)$ for different densities.} Continuous lines are
 exponential laws obtained by fitting the peaks of the correlation function. }
 \label{fig:g6}
\end{figure}

Figure~\ref{fig:g6} shows the hexatic order correlation function $g_6(r)/g(r)$ for different densities. The correlation
length is obtained by fitting the exponential decay of the peaks of the correlation function. We use an exponential function
instead of a 2D Ornstein-Zernike function to avoid \emph{a priori} assumptions on the origin of the growth of the correlation
length.

Positional order is expressed by the pair correlation function, $g(r)$, and its length scale measured by the decay of $g(r)-1$.
The two-body correlation function $\xi_2$ is obtained by fitting with an exponential law the decay of $g(r)-1$.

\subsection*{Estimation of $\xi_{K,random}$ and $\xi_{K,uniform}$}

The static lenghscale $\xi_K$ was measured in both the random and uniform pinning geometries following the procedure defined in Ref.~\cite{charbonneau2013decorrelation}.
First a microscopic overlap function is defined as
$$
w_{mn}(0,t)\equiv\Theta(a-|\mathbf{r}_n(t)-\mathbf{r}_m(0)|),
$$
where $\Theta$ is the Heaviside function, and $a=0.3\sigma$ ensures single occupancy of cells with side $a$.
The overlap function is then averaged both over different equilibrium configurations and for different realization of the pinning
field as
$$
Q_c(t)\equiv\frac{1}{(1-c)N}\langle{\sum_{m,n\notin\mathcal{P}}w_{mn}(0,t)}\rangle,
$$
where $\mathcal{P}$ is the set of $N_p$ pinned particles with concentration $c=N_p/V$, and the brackets stand both
for thermal and disorder averages. The function $Q_c(t)$ is one at $t=0$ and then decays to an equilibrium value that is
a measure of the overlap between the initial configuration and the subspace of configurations that are compatible with the
pinning field. For low concentrations $c$, the function $Q_c(t)$ is expected to decay rapidly, while for high concentrations
$c$ it should decay to a high value. This corresponds to two states: the fluid state with low overlap, and the \emph{glass} state
with high overlap (even in the ideal glass state the overlap is not $Q_c(\infty)=1$ because of thermal fluctuations).
By measuring $Q_c(\infty)-Q_{c=0}(\infty)$ as a function of the concentration $c$, the point-to-set (PTS) correlation length is defined as
$\xi_{PS}=(c_{PS}\rho)^{1/2}$, where $c^*$ is the concentration that locates the transition.
For small system sizes, it was shown that, at low enough temperatures, this transition appears to be a first-order phase transition~\cite{kob2013probing}.
For big system sizes, the barrier which separates the two states become too high to be simulated, so one defines the PTS 
correlation length as the crossover length between small and large overlap. We follow Ref.~\cite{charbonneau2013decorrelation} and
set $Q_{c_{PS}}(\infty)-Q_{c=0}(\infty)\simeq 0.4$. Care has to be taken in ensuring that the simulations are properly
equilibrated, which can be checked by the decay of the self part of the overlap function:
$$
Q_c^s(t)\equiv\frac{1}{(1-c)N}\langle{\sum_{n\notin\mathcal{P}}w_{nn}(0,t)}\rangle.
$$
For polydisperse hard disks with $11\%$ polydispersity, this limits the simulations to the range $\rho\lesssim 0.97$.

\begin{figure}[!t]
 \centering
 \includegraphics[width=8cm]{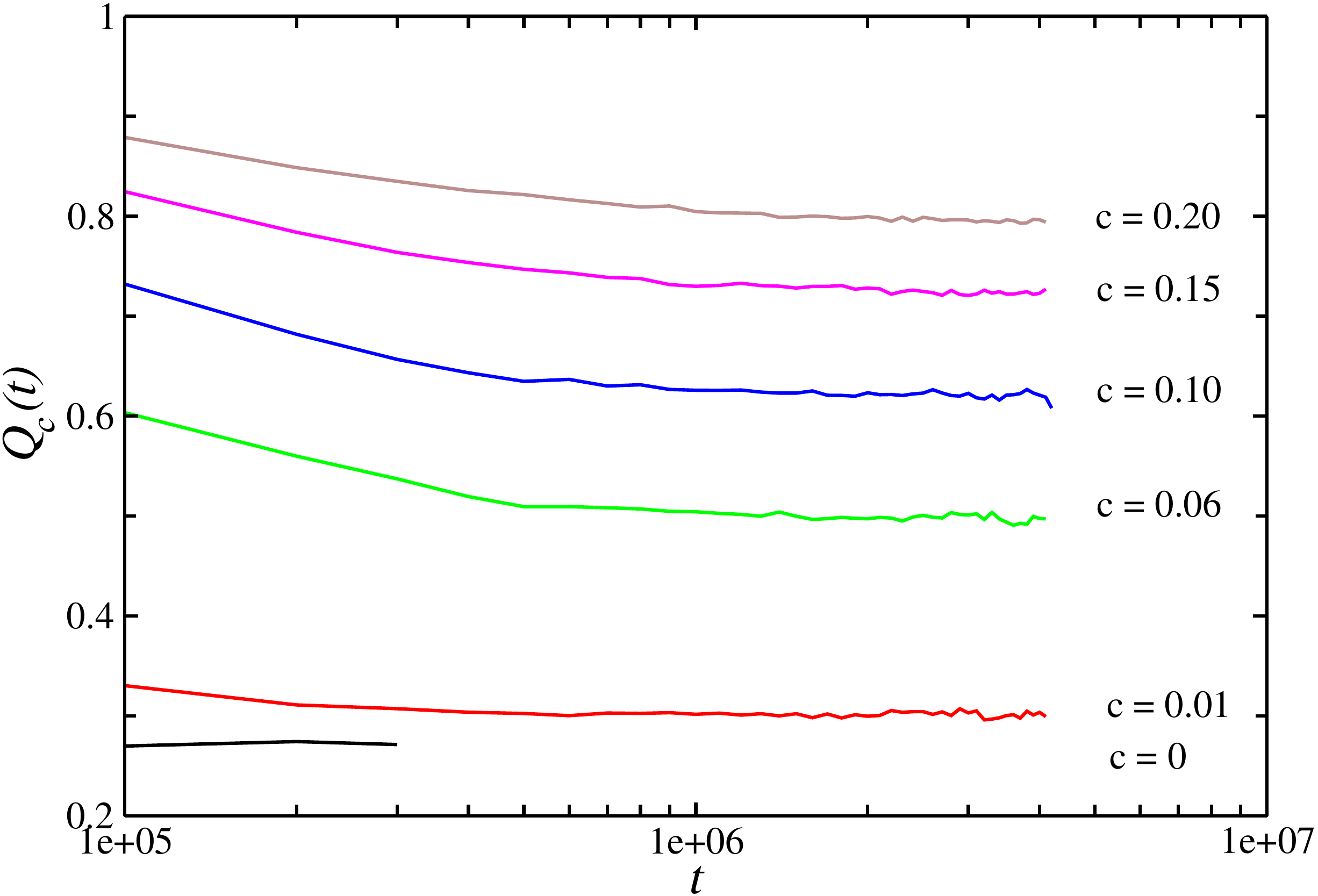}
 \caption{{\bf Average of the overlap function $Q_c(t)$ for different concentrations.} Here $\rho=0.97$.}
 \label{fig:overlap}
\end{figure}

\begin{figure}[!t]
 \centering
 \includegraphics[width=8cm]{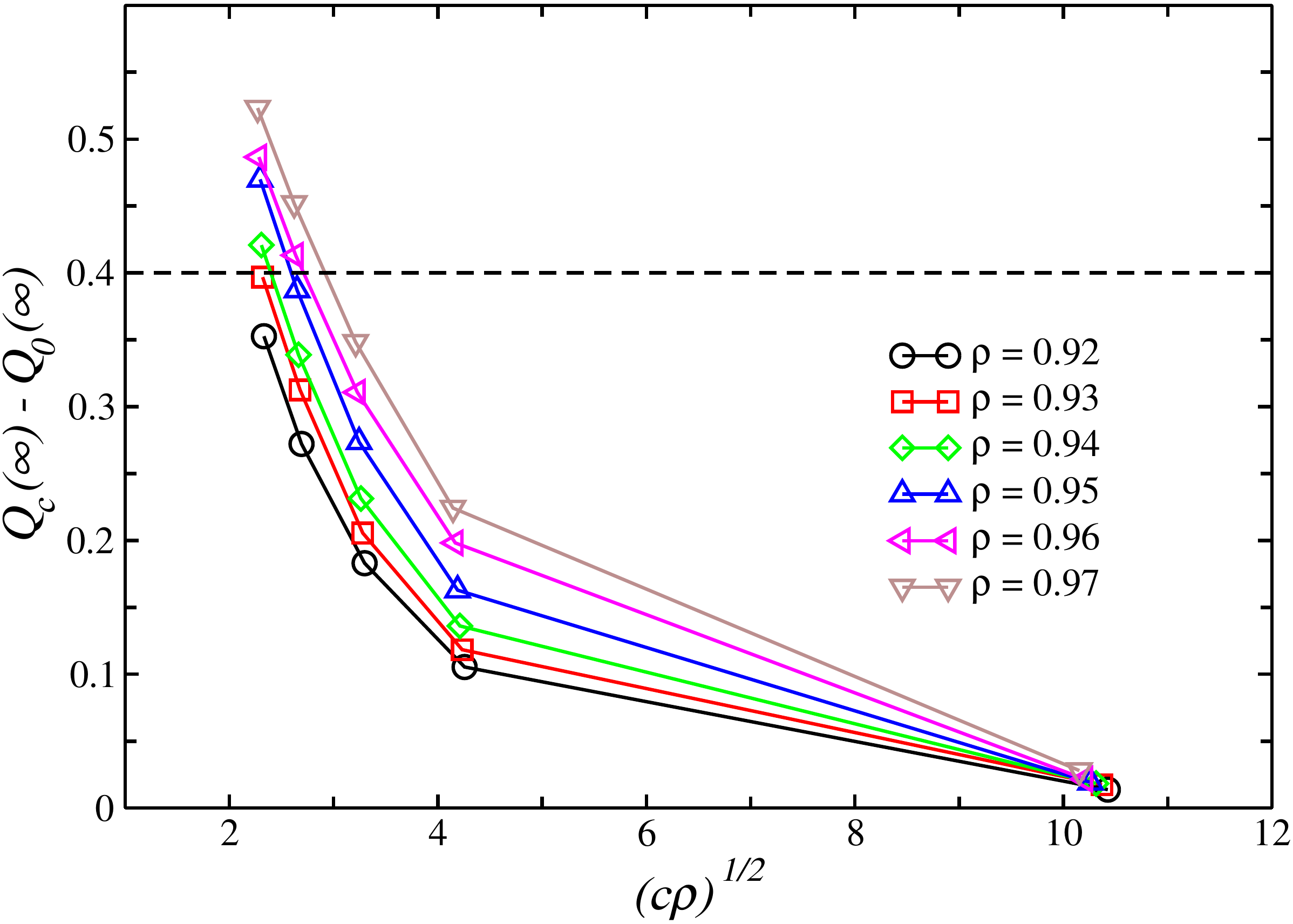}
 \caption{{\bf Overlap as a function of the average distance between the particles for different densities.}
 The PTS correlation length is defined as the length at which $Q_{c_{PS}}(\infty)-Q_{c=0}(\infty)=0.4$.
 For the lower densities this value is obtained by fitting points with $(c\rho)^{-1/2}<10$ by a polynomial and extrapolating
 the result. Given the rapid decay of the overlap, the results do not depend sensibly on the fitting procedure.}
 \label{fig:qcnorm}
\end{figure}

Figure~\ref{fig:overlap} plots the average overlap function $Q_c(t)$
at the highest density considered $\rho=0.97$. The function $Q_c(t)$ decays to a plateau, whose height depends on the concentration of pinned
particles. For low concentrations, the overlap between any two configurations in the system is low, and $Q_c(\infty)-Q_{c=0}(\infty)$ decays to a low value; for high
concentrations of pinned particles, the system is constrained in a region of phase space with high overlap (and overall low configurational
entropy), and $Q_c(\infty)-Q_{c=0}(\infty)$ decays to a high value. The PTS correlation length is defined as the length that characterizes this crossover.
To extract it, we follow Ref.~\cite{charbonneau2013decorrelation} and plot the value of $Q_c(\infty)-Q_{c=0}(\infty)$ versus the average distance between
pinned particles, $(c\rho)^{-1/2}$, as shown in Fig.~\ref{fig:qcnorm}. We fix the crossover value to be around $Q_{c_{K}}(\infty)-Q_{c=0}(\infty)\simeq 0.4$, and define $\xi_{PS}=(c_{K}\rho)^{-1/2}$.
We have checked that the results do not depend sensibly on the exact choice of the crossover value, as the overlap decay rapidly at the crossover.

The same results are confirmed with coarse-grained variables. 
In this approach the simulation box is divided into smaller boxes of side $l=0.3\sigma$, ensuring
that each box can be occupied at most by one disk at any time during the simulation. The problem is then mapped on a set of
discrete variables, defined as $n_i^\alpha=1$ when the center of a disk is in box $i$ in configuration $\alpha$, and $n_i^\alpha=0$
otherwise. The average occupancy for a particular realization of the pinning field is defined as $\overline{n_i}$,
where the overline denotes an average over thermal fluctuations for a fixed realization of the pinning field.

\begin{figure}[!t]
 \centering
 \includegraphics[width=8cm]{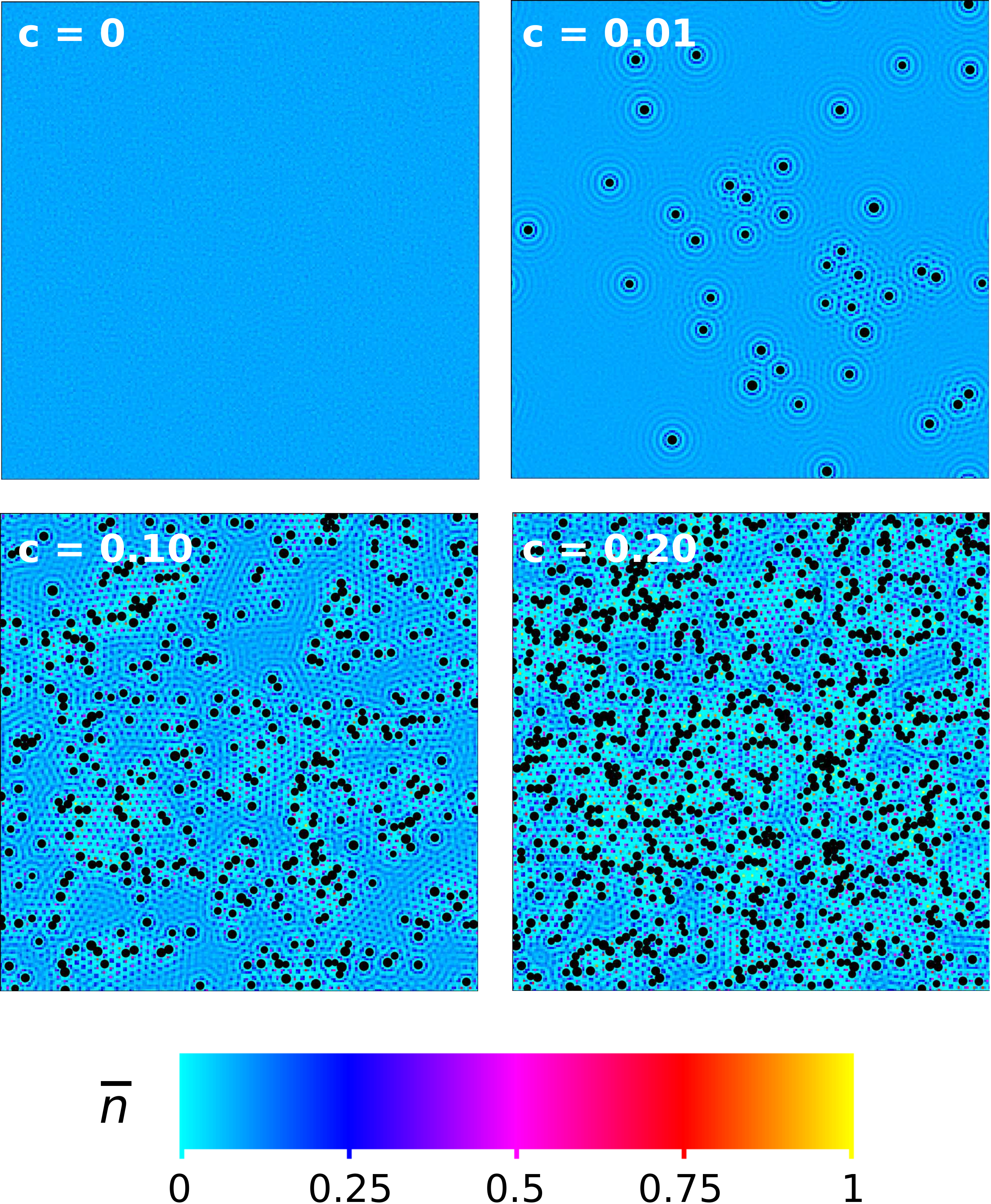}
 \caption{{\bf Average occupancy $\overline{n_i}$ at density $\rho=0.95$.}
 Here we show the $\overline{n}$ field for different concentrations
 of the pinning field, $c=0,0.01,0.10,0.20$.}
 \label{fig:presenceboxes}
\end{figure}

In Fig.~\ref{fig:presenceboxes} we plot $\overline{n_i}$ at density $\rho=0.95$ for concentrations $c=0,0.01,0.10,0.20$.
For zero concentration, the $\overline{n_i}$ is structureless, with a uniform value of $\overline{n_i}\simeq l^2\rho$, where $l$ is the
side length of the boxes. When a pinning field is acting on the system, $\overline{n_i}$ measures the effects of the perturbation. In the figure,
pinned particles are represented by black circles. Each pinned particle generates an oscillatory perturbation of the $\overline{n_i}$ field,
which originates from the two-body static correlations between the pinned particle and the mobile particles in the fluid. This can be seen by measuring the 
spatial correlation function of the $\overline{n_i}$ field,
$$
g_2^{CG}(r_i-r_j)=\langle\overline{n_i}\overline{n_j}\rangle_P-\langle\overline{n_i}\rangle_P\langle\overline{n_j}\rangle_P, 
$$
where the $\langle\cdots\rangle_P$ is an average over different realizations of the pinning field. By fitting the decay
of $g_2^{CG}(r)/g(r)$ with an exponential function, we extract the correlation length $\xi_2^{CG}$.
This length scale is equivalent to the two-body correlation length of the unperturbed system (without pinning field), $\xi_2$.

The coarse grained PTS length is estimated from the following overlap function.
$$
Q_c^{CG}=\langle\langle\frac{1}{N_b}\sum_i^{N_b}n_i^\alpha n_i^\beta\rangle\rangle_P, 
$$
where $N_b$ are boxes that do not contain pinned particles, $\alpha$ and $\beta$ are two arbitrary configurations,
and the brackets denote both thermal and pinned field averages. We evaluate the PTS length of the coarse-grained
field as described for the non-coarse grained variables, with $Q_{c_{PS}}^{CG}(\infty)-Q_{c=0}^{CG}(\infty)=0.25$.
We have extracted the corresponding lengthscale, $\xi_{PS}^{CG}$, and verified that it matches exactly the $\xi_{PTS}$ lengthscale.

The use of coarse graining is not limited to the occupancy variable, $n_i^\alpha$. We can define a coarse grained hexatic field as
$$
\overline{\psi_{6,i}}=\frac{1}{N_T}\sum_\alpha\psi_{6,j}^\alpha\delta_{i,j},
$$
where $N_T$ is the total number of configurations and $\psi_{6,j}^\alpha$ is the hexatic order for particle $j$ in configuration $\alpha$, and
$\delta_{i,j}=1$ if particle $j$ is in box $i$, and zero otherwise. Pinned particles will produce perturbations also in the $\overline{\psi_{6,i}}$
field. By taking the spatial correlation function, $\langle\overline{\psi_{6,i}}^*(r)\overline{\psi_{6,i}}(0)\rangle$, we obtain
the coarse grained correlation length, $\xi_6^{CG}$ which, as expected, is the same length scale
as measured by the bond orientational order in the unperturbed case, $\xi_6$.

\subsection*{Estimation of $\xi_{PTS}$}

The PTS lengthscale $\xi_{\text{PTS}}$ is measured in the cavity geometry following the procedure defined in Ref.~\cite{reichman}.
A coarse grained description of the overlap function is introduced, 
in which space is divided $\tilde{N}$ in cubic boxes of side $a=0.3\sigma$. One then defines the occupancy number $n_i^\alpha$, 
which is 1 if box $i$ in configuration and $\alpha$ is occupied and 0 otherwise.
The coarse-grained overlap is then defined as both thermal and disorder
averages of the following quantity as 
$$
q_R(t)=\frac{1}{a^3\tilde{N}}\sum_i^{\tilde{N}}\langle n_i(t_0)n_i(t_0+t)\rangle. 
$$
The decay of $q_R(t\rightarrow\infty)\equiv q(R)$ is then fitted with a compressed exponential
function as 
$$
q(R)=A\exp{\left(-{\left(\frac{R-a}{\xi_{\text{PS}}}\right)}^\eta\right)},
$$
where $\eta$ is the compressed exponent, and $A$ and $a$ are fitting parameters.

\begin{figure}
 \centering
 \includegraphics[width=6cm,clip]{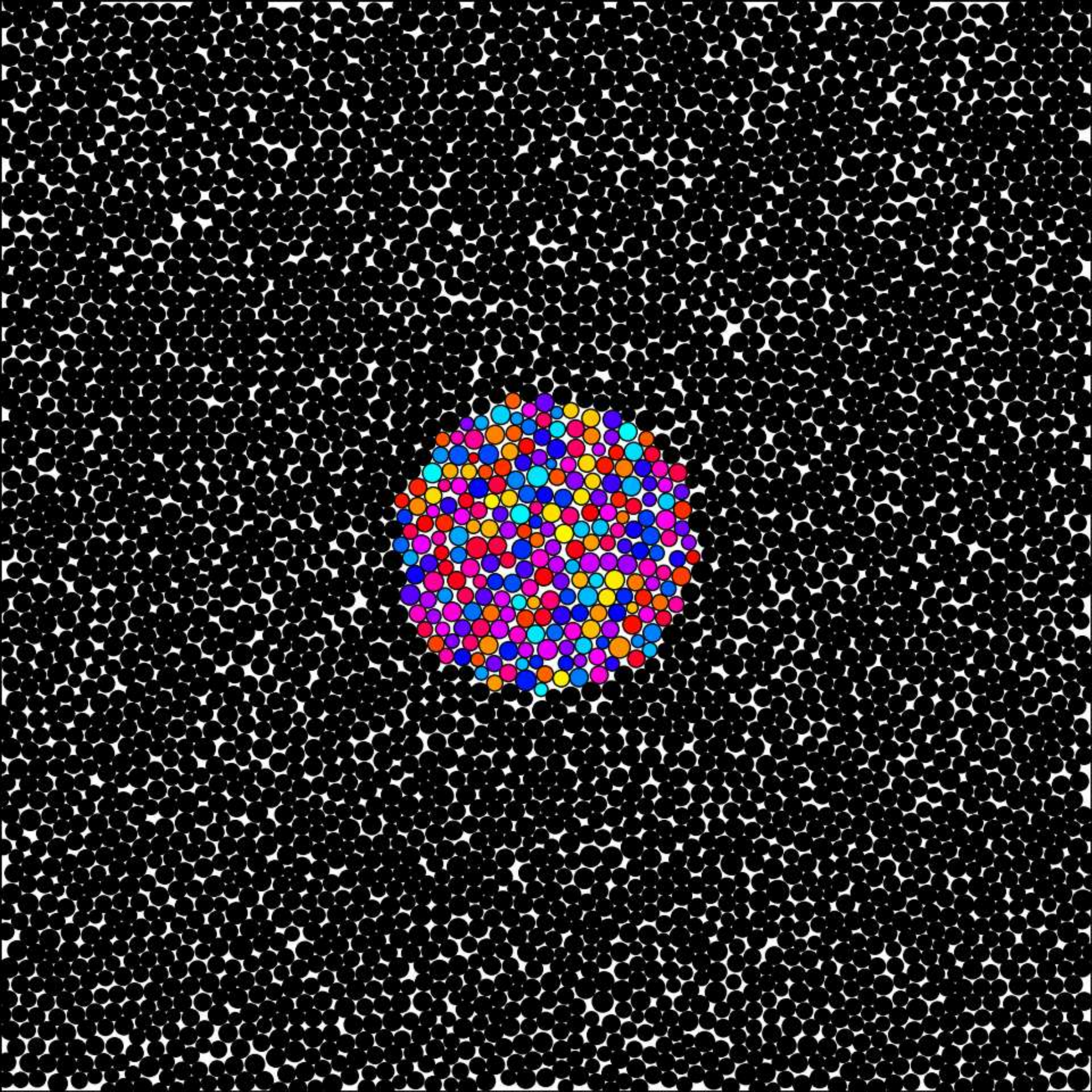}
 \caption{{\bf Snapshot of a configuration at $\rho=0.97$ with a cavity of radius $R=9\sigma$.} Pinned particles are
 coloured in black, while particles inside the cavity are coloured according to their index.}
 \label{fig:cavity_snapshot}
\end{figure}

An example of cavity pinning field is given in Fig.~\ref{fig:cavity_snapshot}. 

In the cavity pinning geometry, the overlap is strongly affected by the location of the cavity in the starting configuration:
cavities inside a region of high hexatic order lead to higher overlaps, while cavities in disordered regions are characterized
by smaller overlaps. A large number of realizations of the pinning field are thus required: in our simulations
we average over ten different initial configurations equilibrated at density $\rho$, and
for each configurations we sample ten cavities in random positions in space, for a total of one hundred simulations
for each density $\rho=0.92,0.4,0.96$ and cavity size $R/\sigma\in [2,9]$, with $\Delta R=\pm 1\sigma$.

\begin{figure}[!h]
 \centering
 \includegraphics[width=8cm,clip]{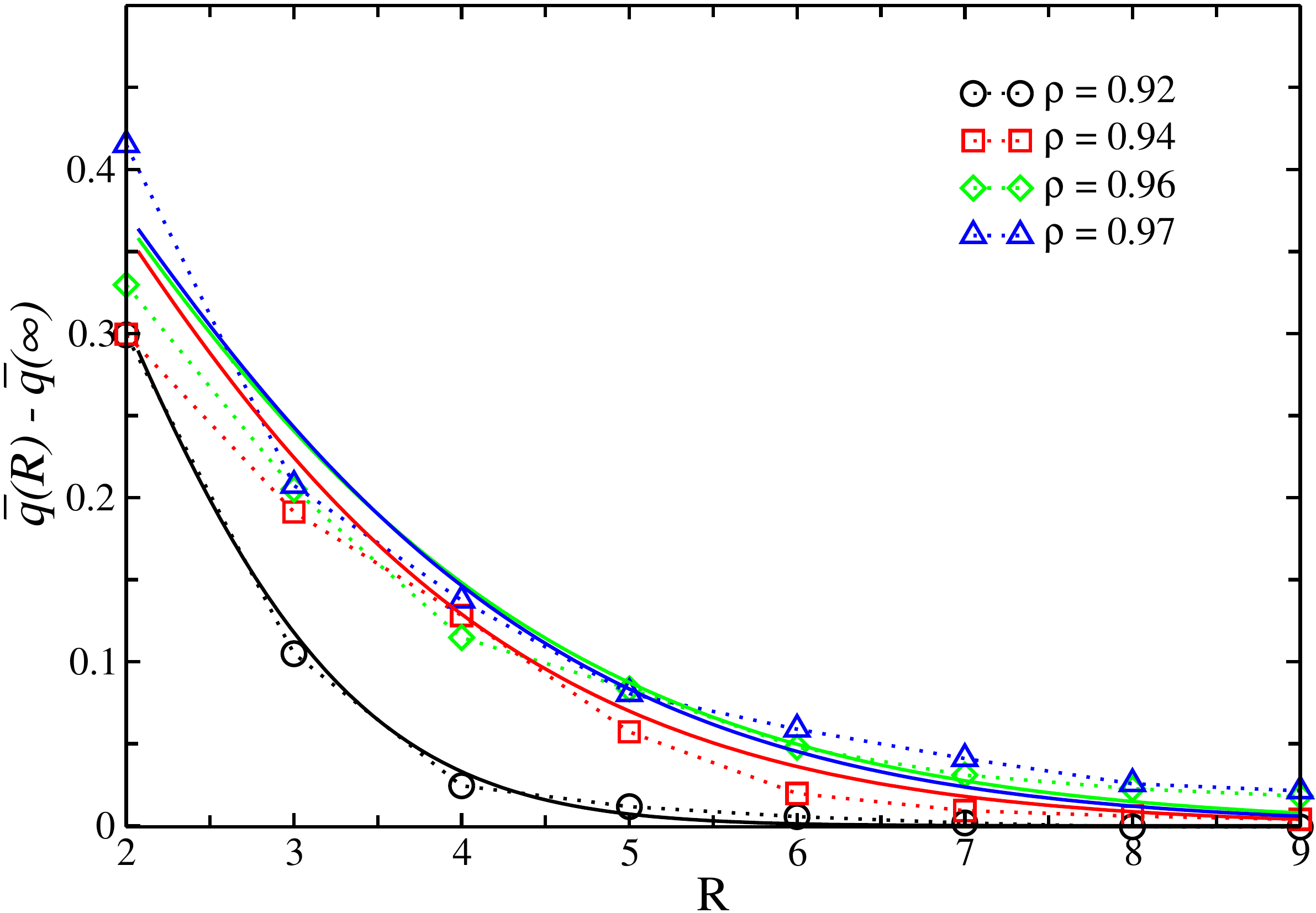}
 \caption{{\bf Overlap (symbols) for cavities of size $R/\sigma\in [2,9]$, with $\Delta R=\pm 1\sigma$.} 
 The results are shown for densities
 $\rho=0.92,0.4,0.96,0.97$. The lines are best fits to a compressed exponential (see text).}
 \label{fig:cavity_overlap}
\end{figure}

From the simulations we extract the PTS length (see Methods) by studying the overlap in 7x7 boxes of side $a=0.3\,\sigma$
located at the center of the cavity. The value of the measured overlap as a function of cavity size $R$ is plotted in Fig.~\ref{fig:cavity_overlap}.

The data in Fig.~\ref{fig:cavity_overlap} were fitted with a compressed exponential: $q(R)=A\exp{\left( -(\frac{R-a}{\xi_{\text{PTS}}})^\eta \right)}$,
with $A=0.5$, following the procedure described in Ref.~\cite{reichman}.

\subsection*{Intermediate scattering function}

\begin{figure}[!b]
 \centering
 \includegraphics[width=8cm,clip]{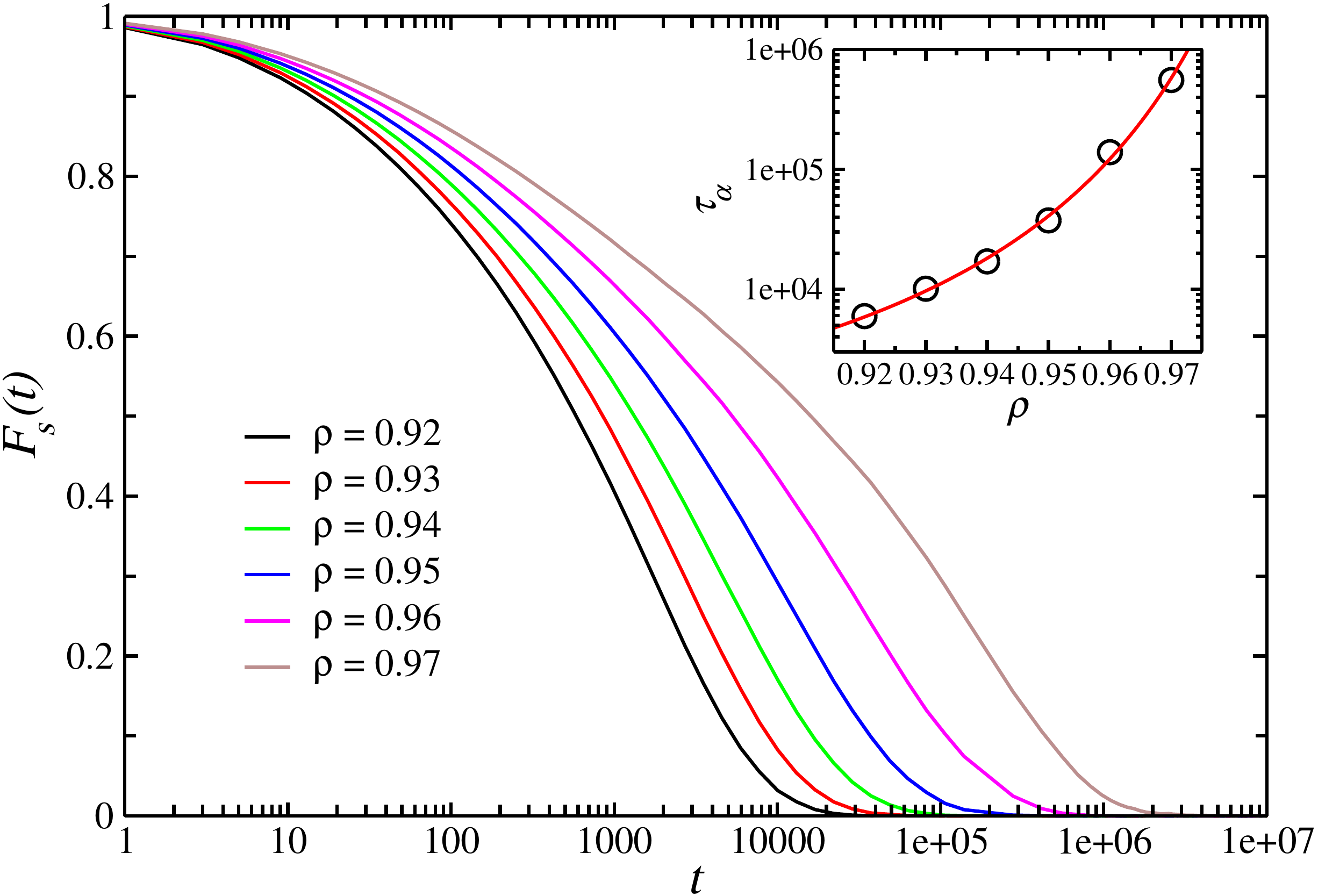}
 \caption{{\bf Density-dependence of self-intermediate scattering function, $F_s(t)$.} 
 It is computed at the wavelength corresponding to the first peak in the structure factor for densities in the range $\rho\in[0.92,0.97]$. 
 The inset shows the structural relaxation time, $\tau_\alpha$ computed as $F_s(\tau_\alpha)=0.1$ (symbols)
 and the VFT fit (line), $\tau_\alpha=A\exp(-D\rho/(\rho-\rho_0))$, with $D=0.42$ and $\rho_0=1.018$.}
 \label{fig:dynamics}
\end{figure}

Figure~\ref{fig:dynamics} plots the intermediate scattering function, $F_s(t)$, for different $\rho$ values, showing
an increase of the relaxation time by more than two orders of magnitude, and a relaxation that acquires
a two-step stretched exponential character. Observing the dynamics at $\rho>0.97$ requires considerably
larger system sizes, as the correlation length of the hexatic order parameter becomes comparable to the simulated system size ($N=10000$ disks)  
(see Fig.~\ref{fig:hexaticorder}). 
The increase of the relaxation time $\tau_\alpha$ (shown in the inset) follows the usual Vogel-Fulcher-Tammann (VFT) law
$\tau_\alpha=A\exp(-D\rho/(\rho-\rho_0))$, where best fits give $D=0.42$ and $\rho_0=1.018$.

\end{document}